\documentclass[10pt,a4paper]{article}

\pdfoutput=1 

\usepackage{graphicx}
\usepackage{float}
\usepackage{amssymb}
\usepackage{amsmath}
\usepackage{datetime}
\usepackage{hyperref}
\usepackage[font=small]{caption}

\usepackage{color}
\usepackage[normalem]{ulem} 
\renewcommand\sout{\bgroup \color[rgb]{0.55,0.00,0.99} \ULdepth=-.5ex \ULset}

\begin{document}

\begin{center}
\begin{figure}
\hspace{1.cm} 
\includegraphics[width=0.1\textwidth]{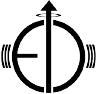}
\hspace{0.5cm}
\includegraphics[width=0.14\textwidth]{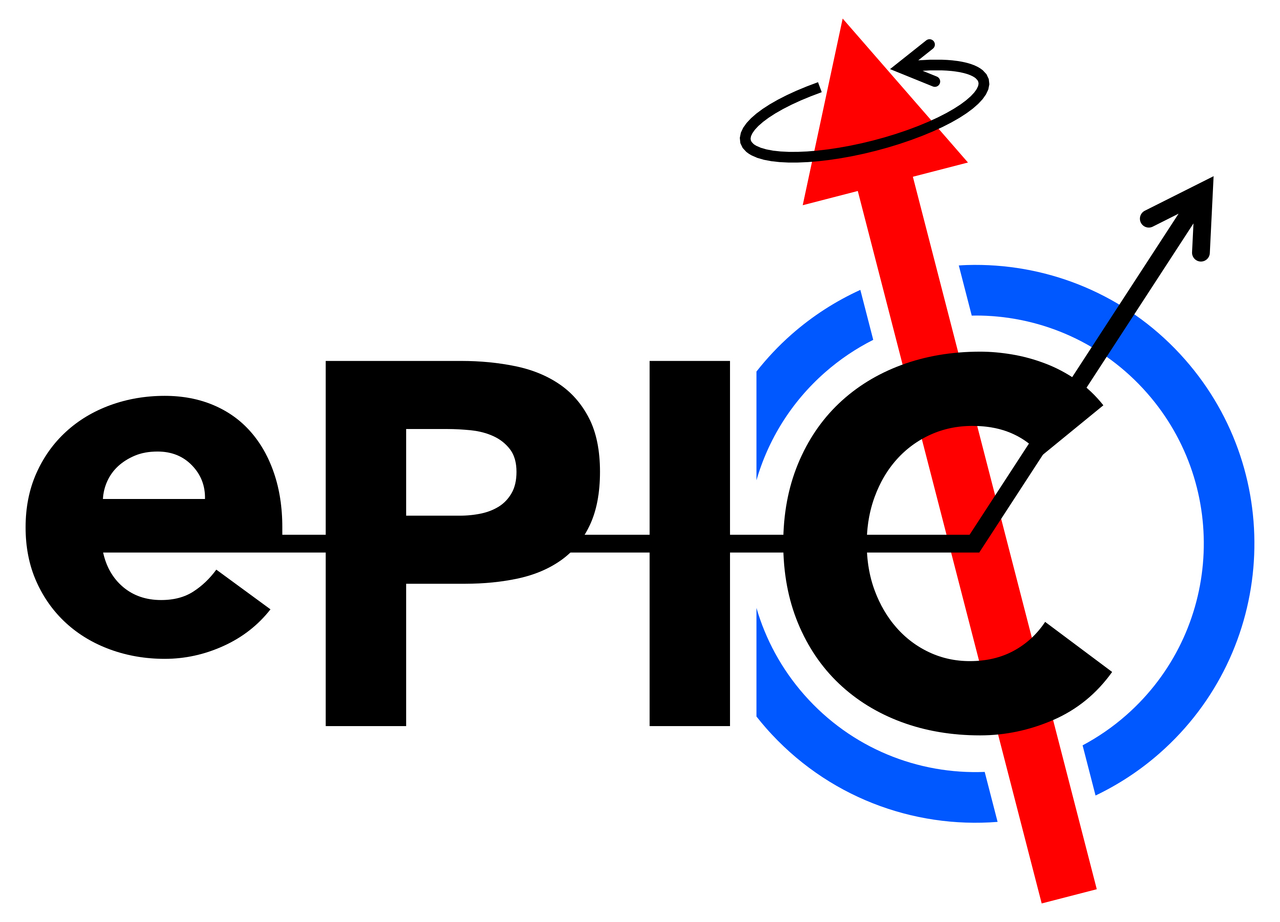}
\end{figure} 
\hspace{10cm} \begin{minipage}{6cm} 
\vspace{-3cm} \begin{center} March 29, 2025 \end{center}
\end{minipage}
\rule[0.5cm]{17cm}{0.2pt}
\end{center}

\begin{center}
{\LARGE \textbf{Synergies between a U.S.-based Electron-Ion Collider}} \\ [0.3cm]
{\LARGE \textbf{and European Research in Particle Physics}}

\vspace{0.7cm}

{\Large Contact Persons: 
    Stefan Diehl$^1$, 
    Raphaël Dupré$^2$,
    Olga Evdokimov$^3$,
    Salvatore Fazio$^4$,
    Ciprian Gal$^5$,
    Tyler Kutz$^6$,
    Rongrong Ma$^{7}$,
    Juliette Mammei$^{8}$,
    Stephen Maple$^{9}$,
    Marco Radici$^{10}$,
    Rosi Reed$^{11}$,
    Ralf Seidl$^{12}$,   
    Zhoudunming Tu$^{13}$}\\[0.3cm]
{\Large On behalf of the ePIC Collaboration and}
{\Large the EIC User Group}




\vspace{1.5cm}

{\large \textbf{Abstract}}

\end{center}
\vspace{0.2cm}

\textnormal{ This document is submitted as input to the European Strategy for Particle Physics Update (ESPPU). The U.S.-based Electron-Ion Collider (EIC) aims at understanding how the complex dynamics of confined quarks and gluons makes up nucleons, nuclei and all visible matter, and determines their macroscopic properties. In April 2024, the EIC project received approval for critical-decision 3A (CD-3A) allowing for Long-Lead Procurement, bringing its realization another step closer. The ePIC Collaboration was established in July 2022 around the realization of a general purpose detector at the EIC. The EIC is based in U.S.A. but is characterized as a genuine international project. In fact, a large group of European scientists is already involved in the EIC community: currently, about a quarter of the EIC User Group (consisting of over 1500 scientists) and 29\% of the ePIC Collaboration (consisting of $\sim$1000 members) is based in Europe. This European involvement is not only an important driver of the EIC, but can also be beneficial to a number of related ongoing and planned particle physics experiments at CERN. In this document, the connections between the scientific questions addressed at CERN and at the EIC are outlined. The aim is to highlight how the many synergies between the CERN Particle Physics research and the EIC project will foster progress at the forefront of collider physics.}

\vspace{2cm}
\noindent
\begin{itemize}


\item[$^1$] {\it 2nd Physics Institute, University of Giessen, Heinrich-Buff-Ring 16,
35392 Giessen, Germany} \\
email: {\tt stefan.diehl@exp2.physik.uni-giessen.de} 

\item[$^2$] {\it Université Paris-Saclay, CNRS, IJCLab, 91405, Orsay, France} \\
email: {\tt raphael.dupre@ijclab.in2p3.fr}

\item[$^3$] {\it University of Illinois Chicago, 1200 West Harrison St., Chicago, Illinois 60607, USA} \\
email: {\tt evdolga@uic.edu}

\item[$^4$] {\it Universit\`a della Calabria \& INFN Cosenza, via P. Bucci, I-87036 Rende (CS), Italy} \\
email: {\tt salvatore.fazio@unical.it}

\item[$^5$] {\it Stony Brook Univerisity of New York, 100 Nicolls Rd, Stony Brook, NY 11794, USA} \\
email: {\tt ciprian@jlab.org}

\item[$^6$] {\it Institut für Kernphysik, Johannes Gutenberg-Universität Mainz, J.J. Becherweg 45, D-55099 \\ Mainz, Germany} \\
email: {\tt tkutz@uni-mainz.de}

\item[$^{7}$] {\it Brookhaven National Laboratory, 98 Rochester St, Upton, NY 11973, USA} \\
email: {\tt marr@bnl.gov}

\item[$^8$] {\it Univerisity of Manitoba, 66 Chancellors Cir, Winnipeg, MB, R3T 2N2, Canada} \\
email: {\tt JMammei@physics.umanitoba.ca}

\item[$^{9}$] {\it University of Birmingham, Edgbaston, Birmingham, B15 2TT, United Kingdom UK} \\
email: {\tt s.maple@bham.ac.uk}

\item[$^{10}$] {\it INFN Sezione di Pavia, via Bassi 6, I-27100 Pavia, Italy} \\
email: {\tt marco.radici@pv.infn.it}

\item[$^{11}$] {\it Lehigh University, 27 Memorial Dr W, Bethlehem, PA 18015, Bethlehem PA, USA} \\
email: {\tt rosijreed@lehigh.edu}

\item[$^{12}$] {\it RIKEN, 2-1 Hirosawa, Wako, Saitama 351-0198, Japan} \\
email: {\tt rseidl@ribf.riken.jp}

\item[$^{13}$] {\it Brookhaven National Laboratory, 98 Rochester St, Upton, NY 11973, USA} \\
email: {\tt zhoudunming@bnl.gov}


\end{itemize}

\newpage


\section{Introduction}
\label{s:intro}

The Electron-Ion Collider (EIC) is a major new research facility to discover and understand the emergent phenomena of Quantum Chromo-Dynamics (QCD). 
Since 1999, there have been many dedicated scientific meetings to shape the physics case for the EIC, most notably the yearly POETIC 
meetings. In 2010 there was a ten-week program at the Institute of Nuclear Theory (INT) in Seattle that resulted in a 547-page document~\cite{Boer:2011fh} detailing the science case. This was updated and condensed in the EIC white paper~\cite{Accardi:2012qut}. 
The U.S. National Academies of Sciences, Engineering, and Medicine (NAS) report~\cite{NAS} summarizes the physics objectives as follows: {\it ``An EIC can uniquely address three profound questions about nucleons - neutrons and protons - and how they are assembled to form the nuclei of atoms:}
\begin{itemize}
\item[-] {\it How does the mass of the nucleon arise?}
\item[-] {\it How does the spin of the nucleon arise?}
\item[-] {\it What are the emergent properties of dense systems of gluons?''}
\end{itemize}
 
To answer these fundamental questions one needs to probe with high resolution and high energy the quark and gluon structure of nucleons and nuclei. It is of great advantage to do this with a simple and well-known probe, such as the electron or the photon. The quark and gluon structure of nucleons is expressed theoretically in terms of parton distributions of various levels of detail and sophistication. For inclusive electron-proton or electron-ion Deep Inelastic Scattering (DIS), where one ignores details of the final state, the scattering process is described in terms of collinear (i.e., transverse-momentum integrated) Parton Distribution Functions (PDFs). When more aspects of the final state are measured, one can become sensitive to the three-dimensional momentum distributions (Transverse Momentum Dependent PDFs or TMDs). Exclusive and diffractive processes allow one to probe also transverse spatial distributions, given in terms of Generalized Parton Distributions (GPDs). The information contained in TMDs and GPDs is truly complementary since the position of partons in a transverse plane is not Fourier conjugate to their transverse momentum. Hence, the combined investigation of PDFs, TMDs, and GPDs, will allow one to arrive at a more complete picture of how nucleons are composed at the level of quarks and gluons. These studies go far beyond global (and scale-dependent) observations like {\it ``50\% of the proton's momentum is carried by gluons''} or {\it ``no more than about 30\% of the proton's spin is carried by quarks''} (a well-known conclusion from the EMC, SMC, and later experiments). More specifically, the three major themes of the EIC physics program that support the NAS physics objectives are:
\begin{itemize}
\item[-] the flavor and spin structure of the proton

\item[-] three-dimensional structure of nucleons and nuclei in momentum and configuration space

\item[-] QCD in nuclei
\end{itemize}

\begin{figure}[ht]
\begin{center}
 \includegraphics[width=0.325\textwidth]{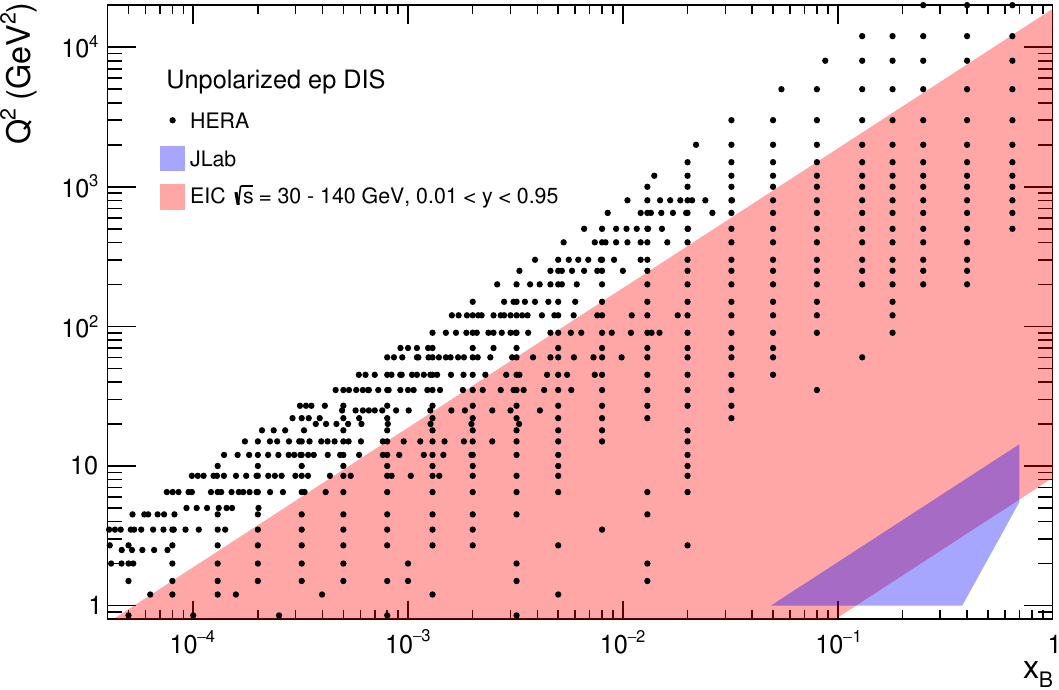}
 \includegraphics[width=0.325\textwidth]{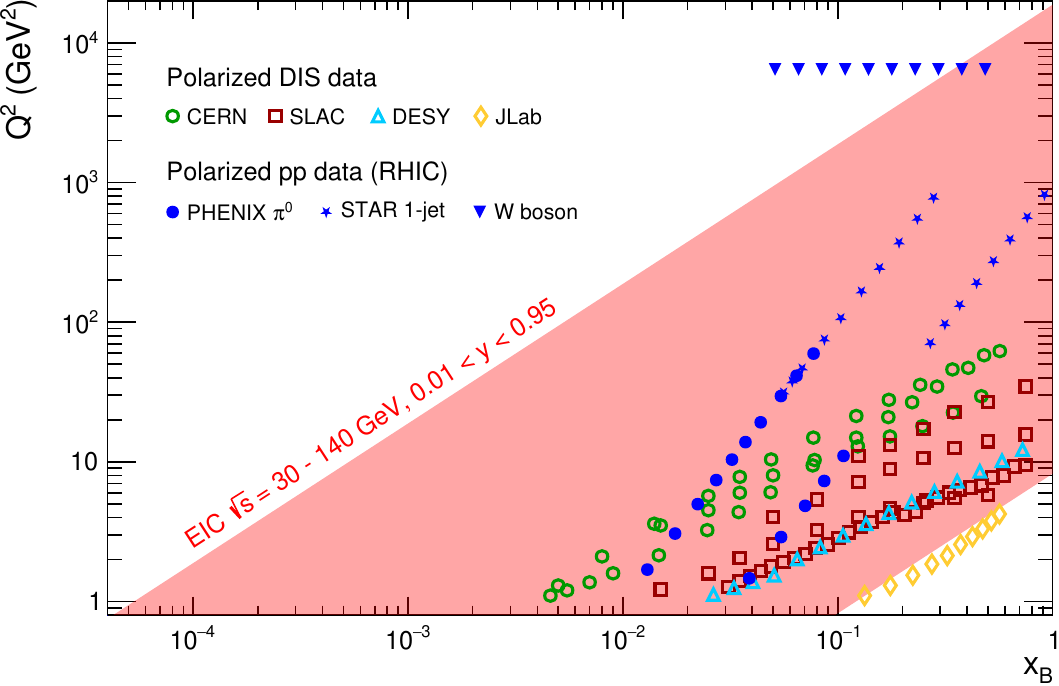}
 \includegraphics[width=0.325\textwidth]{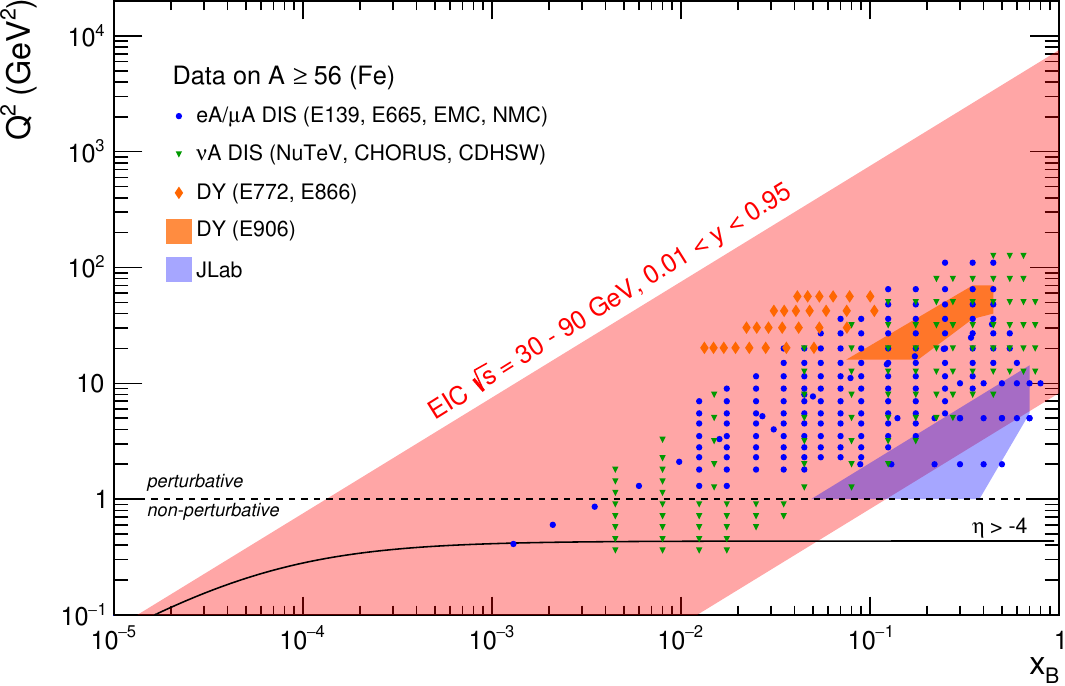}
\begin{minipage}[c]{16 cm}
\usebox0
  \caption{Left and center: the $(x,Q^2)$ phase space coverage of unpolarized and polarized electron-proton DIS for the EIC, in comparison with DIS and heavy-ion data from past and current facilities (updated version of Ref.~\cite{Accardi:2012qut}).  Right: the $(x,Q^2)$ phase space coverage for nuclei, compared to data from existing nuclear DIS experiments.}
  \label{f:fig1}
  \end{minipage}
\end{center}
\end{figure}

The EIC is a machine that is unique compared to any previous collider because of the combined availability of high energy, high luminosity, ion versatility, and polarization.  
It is the first ever machine with the capability to collide highly polarized electrons on polarized protons and light ions, as well as on unpolarized heavier ions up to uranium. The EIC has a large reach in $x$ and $Q^2$ (see Fig.~\ref{f:fig1}). High energy scattering of polarized electrons and ions, including both longitudinally and transversely polarized light ions, is crucial to a full understanding of the quark-gluon structure and dynamics of baryons, mesons, and nuclei. Compared to the HERA collider at DESY, the EIC will have lower energy but a luminosity up to a thousand times higher, enabling measurements that have never been feasible before. The NAS committee found the scientific case for EIC compelling, unique, and timely. According to the NAS report~\cite{NAS}: {\it ``The science questions that an EIC will answer are central to completing an understanding of atoms as well as being integral to the agenda of nuclear physics today. In addition, the development of an EIC would advance accelerator science and technology in nuclear science; it would as well benefit other fields of accelerator-based science and society, from medicine through materials science to elementary particle physics.''} 

In addition, the high luminosity and cleaner environment (with respect to hadron colliders) will enable precision studies in electroweak physics and some specific searches of physics beyond the Standard Model (BSM). 

The fast-growing worldwide community of scientists interested in the EIC organized itself under the EIC Users Group (EICUG) [web site \url{http://www.eicug.org/}]. As of February 2025, the EICUG consists of 1549 scientists (including 377 theorists) from 303 institutions of 40 countries in all world regions, with a large European involvement consisting of 412 scientists (27\%) from 86 institutions. 

In December 2019, following the extremely positive NAS assessment~\cite{NAS}, the US Department of Energy (DoE) established EIC Critical Decision 0 (CD-0), a ``mission need" declaration, formally starting the EIC Project.
Following a call for detector proposals, the ePIC Collaboration was established in July 2022 with the goal of realizing a general purpose detector designed to deliver the full EIC science program. As of February 2025, the ePIC collaboration consists of ~1000 members, of whom 29\% are based in Europe.
In April 2024, the EIC Project received approval for Critical Decision 3A (CD-3A) which allows the EIC project to initiate long lead procurements.

With this document we highlight the elements of the EIC scientific program that will directly or indirectly impact particle physics. In particular, we want to outline to the Panel of the European Strategy for Particle Physics Update (ESPPU) the benefits that will be obtained in combining data from the EIC and CERN experiments. Moreover, this research activity benefits greatly from the significant European involvement in the EICUG and ePIC collaboration as highlighted above. We believe that such 
cross-cutting collaboration is central to the scientific progress in our field and should be strongly encouraged.
 
The general need for and uses of high-energy electron-proton and electron-ion DIS
are outlined in a separate document submitted to this Panel~\cite{ESPPU-DIS}. Furthermore, detailed aspects of the R\&D programs of both the U.S.-based EIC accelerator~\cite{ESPPU-accel} and the ePIC detector~\cite{ESPPU-ePIC} are also outlined in dedicated documents. Here, the focus will be on the EIC physics case, the synergies with present and planned CERN experiments, and the European involvement in the EIC.

The EIC is expected to start operating toward the mid of the 2030's. It will likely run concurrently with LHC after its high-luminosity upgrade. Hence, it seems appropriate and timely to outline below how investigations at the EIC of each of the above topics could benefit CERN experiments and vice versa. 
\noindent


\section{The flavor and spin structure of the proton}
\label{s:spinflavor}

This topic centers around the accurate determination of collinear parton distributions for both unpolarized and polarized protons (and neutrons and deuterons). The unpolarized collinear PDFs currently used for LHC studies 
have an (NNLO) uncertainty of 2.4\% for up quarks at $x=0.5$ and $Q^2=100$ GeV, but 12\% for down quarks, 140\% for strange quarks and 34\% for gluons~\cite{Salam, Rojo}. The uncertainty quickly rises for all flavors at $x \gtrsim 0.6-0.7$ (see Ref.~\cite{Aschenauer:2019kzf}). 

It is important to constrain PDFs in the limit of large $x$ over a wide range of $Q^2$ because they influence the production rate of high transverse momentum $W$ and $Z$ bosons and jets, as well as the possible production of new heavier partners that are predicted in several BSM extensions (see the DIS document for more details). LHC data will help to decrease these uncertainties, particularly after the high-luminosity upgrade~\cite{Khalek:2018mdn}, but for the search for BSM physics at the LHC it is essential that the employed PDFs are obtained from data that are insensitive to that BSM physics. Proposed future experiments at CERN, such as a fixed-target experiment~\cite{Hadjidakis:2018ifr} or the LHeC experiment (see the document on DIS submitted to the ESPPU Panel) could provide such data with high precision. An overall 1\% uncertainty in the PDFs would be the desired goal at the LHC to confront with theory~\cite{Salam}. High statistics data obtained with the EIC from neutral and charged currents in electroweak DIS will help to reach a better precision at large $x$, especially for the EIC configuration with maximum center-of-mass energy of 140 GeV. For instance, a projection study shows that charged current DIS at the EIC would have very strong impact on the $x \bar{d} + x \bar{s}$ combination of quark distributions~\cite{Aschenauer:2019kzf}. Moreover, the EIC will extend these measurements to a completely new domain with effective neutron targets by using deuterium beam and the capability to tag spectator protons~\cite{Jentsch:2021qdp}. This will be a unique dataset to improve our knowledge of the down quark PDF and test isospin symmetry.

The EIC can measure various processes from which the contribution from different quark and antiquark flavors can be separately extracted over a very broad range in $x$. These processes include not only neutral and charged current electroweak DIS but also semi-inclusive DIS (SIDIS), where a hadron in the final state is identified. SIDIS at the EIC is particularly useful for a precise determination of the strange quark distribution~\cite{Aschenauer:2019kzf}. Such analyses go hand in hand with the extraction of collinear fragmentation functions (FFs), where light flavor contributions can be individually tagged. Therefore, our knowledge of PDFs is influenced also by the accuracy at which FFs are known. The FFs are usually determined in $B$-factory experiments by measuring electron-positron collisions, but this yields only a limited knowledge of each individual flavor contribution, and the gluon channel is reachable only at subleading order. Ideally, one should extract FFs by performing a global fit to data of all available reactions, including hadron collider data. At present, this was done only by one group~\cite{deFlorian:2014xna,deFlorian:2017lwf}. More recently, the {\tt NNPDF} collaboration found that hadron collider data ({\tt CDF, CMS, ALICE}) can significantly constrain the gluon FFs~\cite{Bertone:2018ecm}. On the other hand, the (anti)strange FFs still suffer from large uncertainties, and this reflects in a large uncertainty on the (anti)strange PDFs as well. More generally, the limitations of the current SIDIS data hinder a complete flavor separation of unfavored channels. At the EIC, the high luminosity, combined with the large lever arm in the hard scale $Q$ and the purposefully planned detector capabilities, will allow for very precise studies of the flavor dependence of FFs over a large phase space. Current studies on the projected relative error indicate that significant improvements can be achieved also for PDFs of light flavors over a wide range of low to medium $x$, particularly for the strange component~\cite{Aschenauer:2019kzf}. Hence, combining inputs from the EIC, hadron colliders and $B$-factories, will allow to drastically reduce the uncertainties on FFs, and could make it possible also to reach the ultimate goal of a simultaneous extraction from data of both PDFs and FFs. Additionally, it could help in clarifying if the intrinsic flavor content of the proton (i.e., Fock components in its wave function) receives contributions also from charm~\cite{Goncalves:2024elt}. 

\begin{figure}[htbp]
\begin{center}
 \includegraphics[width=0.5\textwidth]{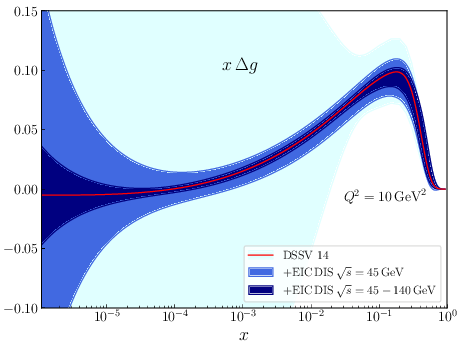}
\begin{minipage}[c]{16 cm}
\usebox0
  \caption{Integrated gluon helicity as a function of the attainable $x_{\mathrm{min}}$ for various EIC configurations~\cite{Borsa:2020lsz}.}
  \label{f:fig2}
  \end{minipage}
  \end{center}
\end{figure}

The availability of polarized proton beams at the EIC allows a similar analysis for the polarized quark and gluon PDFs and thereby the possibility to shed further light on their contribution to the proton spin. In particular, the strange quark and  gluon PDFs still have large uncertainties. Recent results obtained at RHIC give evidence that the gluon contribution is nonvanishing and positive, although the uncertainty is large because the result is very sensitive to the minimum attainable $x$ ($x_{\mathrm{min}}$). With its unique capability of colliding longitudinally polarized electrons and protons, while spanning small $x$ even below $10^{-4}$, the EIC will drastically reduce this uncertainty (see Fig.\ 2). Both DIS and SIDIS data will be important here, as well as the 
development of reliable and versatile Monte Carlo generators to analyze them. 
In this field, Europe has a recognized leadership and can give a crucial contribution. 


\section{Three-dimensional structure of nucleons and nuclei in momentum and configuration space}
\label{s:3dim}

One of the main topics to be studied with the EIC is that of transverse momenta and positions of quarks and gluons inside hadrons, as quantified by TMDs and GPDs, respectively.  The ultimate aim of such studies is to gain a deeper insight into the dynamics of quarks and gluons in hadrons than can be inferred from PDFs alone. TMDs and GPDs have been central to the investigations conducted earlier by the {\tt HERMES} experiment at DESY and currently by the {\tt COMPASS} experiment at CERN and at Jefferson Lab with both polarized and unpolarized targets. Quark TMDs are typically studied using SIDIS and also in jet and vector boson production in proton-proton collisions at {\tt RHIC}. Quark GPDs are typically studied through the Deeply Virtual Compton Scattering (DVCS) and Deeply Virtual Meson Production (DVMP) processes. Polarized light ions at the EIC (both longitudinal and transverse) are crucial for fully resolving the diverse spin-spatial and spin-momentum correlations of hadronic structure. 

\vspace{0.5cm} \noindent
\leftline{{\bf TMD factorization and evolution}}

\noindent
Many critically important aspects of TMDs can be studied experimentally with the EIC. Apart from measuring the various TMDs~\cite{Bacchetta:2006tn} for polarized and unpolarized quarks, gluons in various hadrons, the TMD formalism itself can be thoroughly investigated. 

TMD factorization, which allows for the theoretical description of particle spectra at small or intermediate transverse momentum, needs to be tested. A main objective is to demonstrate to what extent the TMDs are universal (like the collinear PDFs and FFs). Although the TMD for unpolarized quarks and hadrons is expected to be the same in SIDIS and Drell-Yan (DY), for some polarized cases there is a calculable process dependence. In 2015 and 2018 the {\tt COMPASS} experiment acquired DY data with the objective of testing this (more specifically, determining the predicted overall sign difference between the SIDIS and DY measurements of the Sivers effect) and the results so far are consistent with the expectations from the TMD formalism, although the uncertainties are still large. The same applies to the asymmetry measurements at RHIC in $W$ production~\cite{STAR16}. Furthermore, for gluon TMDs even the unpolarized case is expected to be non-universal~\cite{Buffing:2013kca}. Hence, it is important to compare observables 
at the EIC with related observables at the LHC, 
e.g., for quarkonium production~\cite{BoerPisano}
or Higgs production~\cite{BoerDenDunnen}, where it was shown that the transverse momentum distribution of bosons as heavy as the Higgs particle can be affected by TMDs of gluons with small intrinsic transverse momentum. 
The specific predictions from TMD formalism can be safely tested in a high $Q^2$ regime $(Q^2 \gg M_p^2)$, in order to avoid complications from power suppressed quark-gluon correlation effects. In addition, TMD evolution, i.e., the $Q^2$ dependence of the predictions, needs to be tested, as was done extensively for the DIS scaling violations at HERA, showing that Altarelli-Parisi equations (DGLAP evolution of collinear PDFs) work over many orders of magnitude, but at the same time hinting at possible deviations at small $x$ and moderate $Q^2$~\cite{Ball18}. Investigation of the latter regime is another objective of the EIC (covered in Sec.~\ref{s:nuclei}). 

\vspace{0.5cm} \noindent
\leftline{{\bf Transition between TMD and collinear frameworks}}

\noindent
Besides testing its predictions, the limitations of the TMD formalism also need to be clarified. For this purpose proton-proton collisions, even unpolarized ones, are particularly useful. Currently, it is expected on theoretical grounds that azimuthal correlations in dijet production in $p$-$p$ collisions do not factorize~\cite{factbreak}. The size of the factorization breaking effects is entirely unknown and can only be assessed from a comparison of high energy $p$-$p$ and $e$-$p$ collisions. The existing {\tt HERA} data on dijet production in $e$-$p$ collisions is too limited to do this. Recent studies indicate that dijet measurements at the EIC are feasible~\cite{Dumitru:2018kuw,Zheng:2018awe}. The same applies to $D$-meson pair production, which is another process of interest for tests of TMD factorization breaking. 

The TMD formalism applies to the region of low transverse momenta (much smaller than the large scale in the process). Therefore, sufficient momentum resolution is required in the full $p_T$ range in order to test the limits of applicability and the transition to the collinear formalism that applies at large $p_T$. In many of the observables of interest, such as azimuthal correlations in the Drell-Yan process (studied by, e.g., {\tt CMS} \cite{Khachatryan:2015paa} and {\tt ATLAS} \cite{Aad:2016izn}), the momentum resolution is currently limited, but that will improve over time. The same applies to di-photon and quarkonium (pair) production. The transition region of intermediate $p_T$ has recently attracted much attention from the theoretical point of view~\cite{Collins:2016hqq,Echevarria:2018qyi}. Because of the additional degrees of freedom (observables differential in more variables) it is harder to reach the same level of precision as for observables calculable in collinear factorization. Nevertheless, at present  expressions at next-to-next-to-next-to-leading-logarithmic accuracy (N$^3$LL) are available for a number of processes{\color{blue}~\cite{MAP,Art23}}. More data, from LHC and other experiments, are needed to confront with theoretical predictions. 

\vspace{0.5cm} \noindent
\leftline{{\bf GPDs}}

\noindent
GPDs enter collinear factorization expressions and are expected to be universal. While the associated theory has reached a high level of sophistication, progress on relevant measurements has been comparatively slow because the experiments require high statistics for exclusive processes and excellent detector coverage. 
{\tt HERA, COMPASS} and Jefferson Lab experiments have already obtained experimental information on some of the GPDs, but this is only the beginning when it comes to extractions of the GPDs with precision and in a sufficiently large kinematic domain. The capabilities of the EIC will open the way to a thorough exploration of GPD properties. In particular, the large $Q^2$ coverage will enable tests of GPD evolution and studies of power-suppressed higher-order correlations. 

GPDs encode information about the spatial distribution of partons inside a hadron, correlated with their distribution in longitudinal momentum. The spatial distribution is obtained in a rather direct way by Fourier transforming the differential cross section as a function of the Mandelstamm variable, $t$, 
in suitable exclusive reactions like DVCS or DVMP. In an indirect manner, this distribution also influences the dynamics of $p$-$p$ as well as $p$-$A$ and $A$-$A$ collisions, namely in the context of multiparton interactions (MPI). In such interactions, several partons in the colliding hadrons take place in independent hard scatters, and the relative transverse distance between the partons is of crucial importance for this mechanism. Information on the spatial distribution of single partons from GPDs provides a quantitative baseline expectation, on top of which one can then attempt to assess correlation effects between different partons. In this sense, MPI and underlying events description will considerably benefit from a fully developed GPD picture. 

Moreover, GPDs will enable detailed studies of the spatial distribution (tomography) of several interesting observables like charge, pressure, energy and number densities. 

\vspace{0.5cm}

\noindent
TMDs and GPDs have been studied extensively. Various computer codes and tools, to a large extent developed by European groups, are available, such as for example: 
\newline {\tt TMDlib} and {\tt TMDplotter} (\url{https://tmdlib.hepforge.org/doxy/html/index.html/}), 
\newline {\tt NangaParbat} (\url{https://github.com/MapCollaboration/NangaParbat}), 
\newline {\tt Artemide} (\url{https://github.com/VladimirovAlexey/artemide-public}). 
\newline For GPD model studies also the package {\tt PARTONS} (\url{https://arxiv.org/abs/1512.06174}), which is the basis for EpIC~\cite{Aschenauer:2022aeb}, a newly developed generator of exclusive processes. 
\newline There are also proposals for future TMD and GPD studies at CERN. 
The AMBER experiment at the M2 beam line of the CERN SPS started operation in 2023~\cite{AMBERprop}, planning Drell-Yan measurements with pion and kaon beams.
Another initiative that is being put forward is that of a fixed-target program which would allow scattering of an LHC proton or lead beam on a polarized fixed-target.
The physics case for such an experiment has been developed over recent years under the name ``AFTER@LHC''~\cite{Hadjidakis:2018ifr}; a particular proposal for such a fixed-target experiment  at {\tt LHCb} is called ``LHC-spin'' and is described in a separate document submitted to this Panel~\cite{ESPPU-LHCspin}. There would of course be ample synergies between such experiments and the EIC as well.


\section{QCD in nuclei}
\label{s:nuclei}

The EIC will be the first collider ever for deeply inelastic scattering with nuclei, opening up a large new phace space in high atomic number $A$ at small $x$ and large $Q^2$ that has never been experimentally accessed (see Fig.~\ref{f:fig1}). Electron-ion collisions allow the study of the internal structure of heavy ions in terms of elementary partonic constituents, quarks and gluons. In addition to a new understanding of QCD in large nuclei being a fundamental topic in itself, it is also complementary to the program of heavy-ion collisions at the LHC and RHIC. Probing the partonic structure of the colliding nuclei by DIS experiments is important for understanding the production of both the matter that then becomes a quark-gluon plasma, and the jets and other hard probes that are used to explore its properties. 

\vspace{0.5cm} \noindent
\leftline{{\bf Nuclear PDFs}}

\noindent
At LHC kinematics, one is mostly sensitive to the region of relatively small $x$, where the nuclear PDFs are very poorly known since they are hardly constrained by any currently existing data. Studies of modifications to the structure of jets, as they pass through deconfined QCD matter, are becoming very sophisticated and will form an increasingly important part of the nuclear collision program at the LHC, when it moves to higher luminosities. These studies will require a reduction of the large uncertainty of nuclear PDFs at small $x$, and the EIC will have a significant impact here~\cite{Aschenauer:2017oxs}. In particular, one should note that accurate measurements of charm structure functions at the EIC would have an even larger effect than total cross section measurements, also improving the determination of the nuclear gluon PDF at large $x$ (aiding the further study of the EMC effect and of (anti-) shadowing). SIDIS at large $x$ in nuclei also provides a laboratory to study the energy loss of high-energy partons passing through ordinary confined nuclear matter. Such measurements provide the comparison that is necessary to calibrate and check theoretical approaches to jet energy loss. Last but not least, exclusive and semi-inclusive measurements will give access to the GPDs and TMDs of nuclei, respectively. 

\vspace{0.5cm} \noindent
\leftline{{\bf Saturation phenomena}}

\noindent
At high collision energies, or equivalently small $x$, the phase space available for emitting soft gluons is very large. Since every emitted gluon is itself a source of further radiation, a fast growing cascade is created that leads to a strong growth of the gluon distribution. Eventually unitarity is expected to be preserved. Due to nonlinear interactions, gluon merging starts to play an important role leading to the phenomenon of gluon saturation. Gluon saturation then controls the physics in DIS collisions in the small $x$ limit at moderate $Q^2$. The gluons in the same regime are also responsible for the production of deconfined QCD matter -- the quark-gluon plasma -- in heavy ion collisions, studied at the LHC and in particular by the {\tt ALICE} collaboration. 

Since a high atomic number $A$ also increases the gluon density at a given $Q^2$, gluon saturation is accessible in DIS experiments at lower energies with nuclei than with protons. The EIC will provide a versatile experimental program to investigate in detail this new regime of QCD. In order to fully understand this regime, it is important to simultaneously measure inclusive and semi-inclusive cross sections, inclusive diffraction (diffractive structure functions) and exclusive reactions, such as vector meson production and diffractive dijets. The EIC is being designed to be a facility that can perform this broad set of measurements that are necessary for a full picture of the gluonic structure of nuclei.

The small-$x$ physics at the EIC is complemented by several measurements that are performed in hadron colliders (RHIC and LHC) to probe aspects of the same physics. Inclusive particle production and two-particle correlations in $p$-$p$ and $p$-$A$ collisions, especially at forward rapidities, directly probe the small-$x$ gluons in nuclei. At CERN, these forward measurements are performed at {\tt LHCb} and, especially with future instrumentation upgrades, at {\tt CMS} and {\tt ATLAS}. Furthermore, as already mentioned the small-$x$ and high-$A$ program at the EIC is very closely connected to studies of deconfined QCD matter in heavy-ion collisions. The correct interpretation of several aspects of heavy-ion collisions, such as the so-called "ridge", requires data from high-energy electron-ion experiments. 

\vspace{0.5cm} \noindent
\leftline{{\bf Initial conditions for Quark-Gluon Plasma studies}}

\noindent
One of the most challenging problems in heavy-ion physics is to understand how the gluons and quarks from the colliding nuclei form a thermalized plasma. Data from $A$-$A$ collisions seems to be well described by models assuming a very quick formation of an equilibrated medium. It is, however, very difficult to get direct experimental access to the earliest stages of a heavy-ion collision, and the theoretical understanding of the thermalization process is still quite incomplete. Here, the picture of the small-$x$ degrees of freedom in the nucleus obtained from electron-ion collisions is crucial, as it provides the starting point -- the initial condition -- from which one evolves towards a de-confined matter. 

In recent studies at the LHC, it has become clear that effects usually attributed to collective behavior in $A$-$A$ collisions, such as elliptic flow and the ridge, are also visible in $p$-$A$ and $p$-$p$ collisions. There is currently an intense debate in the field concerning the correct interpretation of these results. They have been explained in terms of multi-particle correlations either already present in the colliding protons and  nuclei or, alternatively, generated by collective interactions when QCD matter is deconfined. For a resolution of the puzzle posed by these results, a baseline measurement of such initial-state correlations in a more tractable system is essential. In this case, the ideal experiments are provided by $e$-$p$ and $e$-$A$ collisions.

\newpage
\noindent
\leftline{{\bf UPCs}}

\noindent
Besides being a source for many small-$x$ gluons, ultra-relativistic heavy ions also form a source of strong electromagnetic fields, which can be probed in Ultra-Peripheral Collisions (UPCs) at high energies. This effectively leads to high-energy photon-ion scattering, where partons at small $x$ are probed both in nuclei (in $A$-$A$ collisions) and protons (in $p$-$A$ collisions). Since in these collisions the photon is always quasi-real, one does  not have the same ability to vary $Q^2$ as in DIS experiments, but on the other hand the higher collision energy gives access to smaller values of $x$ than available at the EIC. To date, exclusive vector meson production in $\gamma$-$A$ collisions measured by {\tt ALICE}~\cite{Abbas:2013oua} have been used to probe nuclear gluons at small $x$. The possibility to separately perform coherent (nucleus stays intact) and incoherent (nucleus breaks up into smaller color neutral fragments) measurements gives an additional handle on probing the nuclear geometry and its fluctuations, which are important features for understanding the initial state of $A$-$A$ collisions. Exclusive $J/\psi$ and $\Upsilon$ productions in UPCs have been studied at {\tt ALICE, LHCb} and {\tt CMS}. The results (e.g., see Refs.~\cite{TheALICE:2014dwa,Aaij:2015kea,CMS18}) are consistent with {\tt HERA} measurements of the same process at lower energies, and can be used to constrain the $x$ dependence in different theory calculations.


\section{Electroweak physics and the search for physics beyond the Standard Model}
\label{s:ewbsm}

\begin{figure}[htbp]
\begin{center}
 \includegraphics[width=0.6\textwidth]{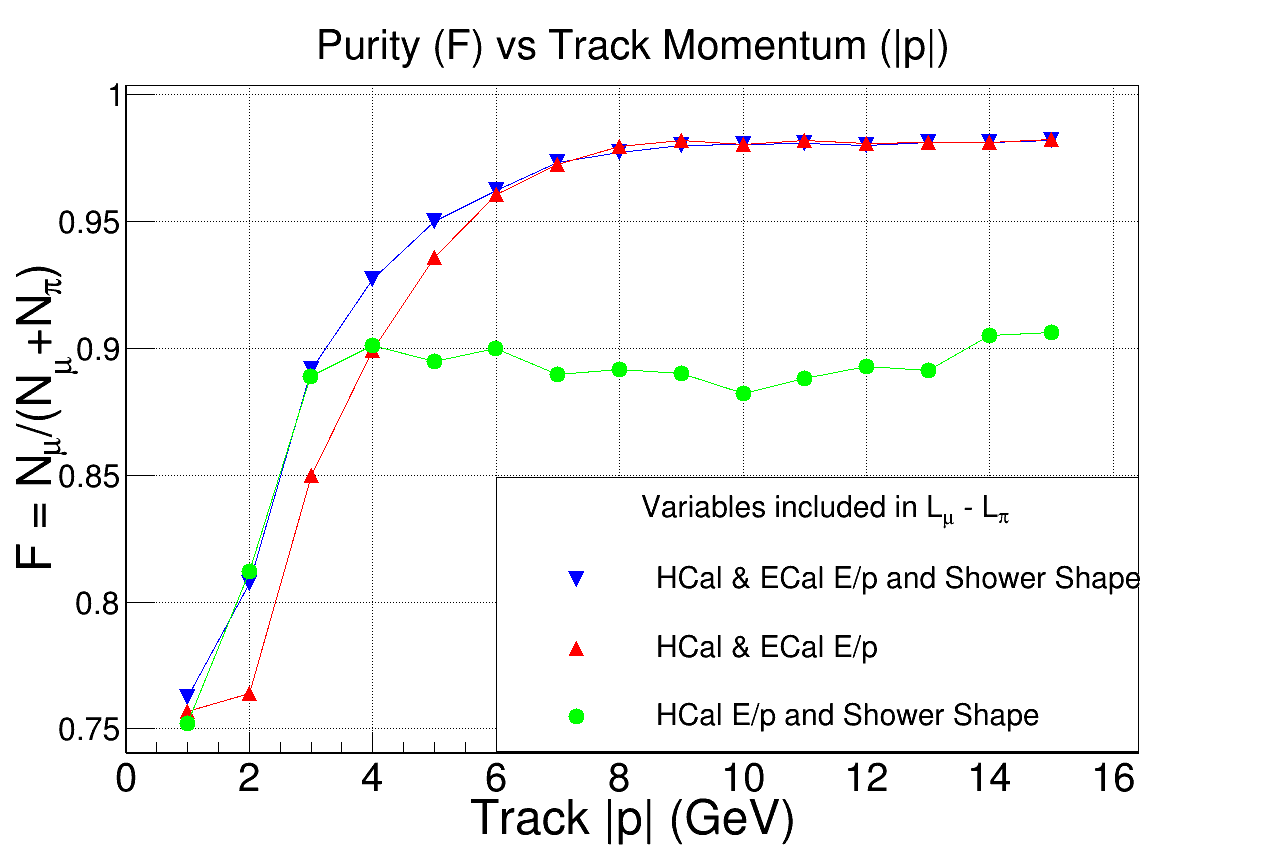}
\begin{minipage}[c]{16 cm}
\usebox0
  \caption{Figure of merit for muon identification in the ePIC detector as a function of momentum.}
  \label{fig:muID}
  \end{minipage}
  \end{center}
\end{figure}

The several TeV limits on new particles established by the LHC indicate that if new BSM physics exists it would reside at a significantly higher mass than the SM. In the absence of a much higher energy collider the ideal way to search for new physics is through precision measurements. These measurements could detect the impact of new particles as they may provide corrections through higher order diagrams. A framework called SMEFT~\cite{Boughezal:2022pmb} has been established to evaluate such effects and has shown that the EIC can have a unique capability to place constraints on new physics. In particular, the EIC polarization for both electrons and protons or deuterons, coupled with its high luminosity, constrains Wilson coefficients that are orthogonal to those probed by Drell-Yan semi-leptonic four fermion interactions at the high luminosity LHC. Additionally, LHC constraints are going to be improved through the increase of knowledge of proton PDFs. 

A second important constraint on BSM physics, that the EIC can provide, is in the sector of charged lepton flavor violation (CLFV)~\cite{eicug_2014_6423359,Banerjee:2022xuw}. In this case, the relatively clean collider environment provided by high energy DIS scattering could easily disentangle an electron to $\tau$ transition. Several measurements have been proposed looking for $\tau$ decays in the final state. The kinematically clean 3-pion decay has been well developed~\cite{Zhang:2022zuz}. This channel alone has been shown to be competitive with other constraints from other facilities around the world. Recent analyses have shown that although the ePIC detector does not have a dedicated muon detector, it can still differentiate muons from pions with a high degree of confidence (see figure~\ref{fig:muID}). This opens up new avenues of investigation, especially in the $\tau$ to single muon decay that has twice as large a branching ratio compared to the 3-pion decay.

Besides the constraints offered by electron-proton or electron-deuteron collisions, the EIC can provide additional information for BSM physics from electron-nucleus scattering. One such study~\cite{Davoudiasl:2021mjy} focuses on coherent production of $\tau$ leptons mediated by an Axion-like particle (ALP). Even with conservative estimates, the authors have shown that the EIC can put significant constraints on ALPs at or above the GeV scale. Crucially, the analysis depends on detection capabilities not currently covered by the baseline ePIC detector, leading to opportunities for an upgrade or a future second detector to be able to explore this new physics.

Finally, the search for new BSM physics relies on the best possible determination of the SM parameters. Among them, the $W$ mass can be precisely extracted also by fitting the transverse mass and momentum distributions of the $W$ decay products in hadronic collisions~\cite{Aaboud:2017svj,Aaltonen:2013vwa}. The ability of making precise extractions of TMDs at the EIC, including sensitivity of intrinsic transverse momenta to the flavor of partons entering the collision, might lead to a statistically significant impact of this nonperturbative effect on the extracted values of $W^\pm$ masses (see Ref.~\cite{Bacchetta:2018lna} for an exploratory work), thus influencing also the search for new BSM physics.

\section{Conclusions}
\label{s:end}

In this document for the ESPPU, we have outlined the European involvement in the U.S.-based Electron-Ion Collider (EIC) which recently received approval for CD-3A allowing the Project to start long-lead procurements. More than a quarter of the EIC User Group and of the ePIC experimental collaboration consists of European scientists. This indicates a large European interest in the EIC, in both its science case and its detector and accelerator developments. Furthermore, this document reviews the large variety of mutual benefits for CERN and EIC experiments coming from the connections between the scientific questions addressed. We conclude that strengthening the ties between the particle physics community in Europe and the EIC project would be very beneficial to all parties involved and could foster important progress in research at the forefront of collider physics.







\end{document}